\documentclass[pre,onecolumn,showpacs]{revtex4}
\usepackage{epsfig}
\usepackage{graphicx}
\usepackage{graphics}

\graphicspath{{figs/}}

\begin{document}

\title{The transition to irreversibility in sheared suspensions: An analysis based on a mesoscopic entropy production}


\author{I. Santamar\'{i}a-Holek$^\dag$, G. Barrios del Valle$^\dag$, J. M. Rubi$^*$}
\affiliation{$^\dag$Facultad de Ciencias, Universidad Nacional
Aut\'{o}noma de
M\'{e}xico.\\
Circuito exterior de Ciudad Universitaria. 04510, D. F.,
M\'{e}xico.\\}
\affiliation{$^*$ Facultat de F\'{\i}sica, Universitat de Barcelona. \\
Av. Diagonal 647, 08028, Barcelona, Spain.}

\begin{abstract}

We study the shear-induced diffusion effect and the transition to
irreversibility in suspensions under oscillatory shear flow by
performing an analysis of the entropy production associated to the
motion of the particles. We show that the Onsager coupling between
different contributions to the entropy production is responsible
for the scaling of the mean square displacement on particle
diameter and applied strain. We also show that the shear-induced
effective diffusion coefficient depends on the volume fraction and
use Lattice-Boltzmann simulations to characterize the effect
through the power spectrum of particle positions for different
Reynolds numbers and volume fractions. Our study gives a
thermodynamic explanation of the the transition to irreversibility
through a pertinent analysis of the second law of thermodynamics.
\end{abstract}

 \maketitle

\section{INTRODUCTION}

When a suspension of non-Brownian particles is subjected to an
oscillatory shear flow, the dynamics of the particles presents a
transition to irreversibility which has been recently observed in
experiments~\cite{pine,acrivos,bredveld}. In these experiments,
the suspension of polymethylmethacrylate (PMMA) particles having
sufficiently large sizes (diameter $d \simeq 230\mu m$), is
contained in a cylindrical Couette cell and taken out of
equilibrium by applying an oscillating shear flow proportional to
$\dot{\gamma}\,cos(\omega t)$, where $\dot{\gamma}=\omega
\gamma_0$ with $\gamma_0$ the applied strain and $\omega$ the
characteristic frequency of the oscillation. At small enough
Reynolds numbers it is observed that the motion of the particles
is oscillatory and reversible, according to a classical result of
hydrodynamics~\cite{taylor}. When increasing the Reynolds number
or the concentration of particles, the trajectories of the
particles become chaotic and then their reversible behavior is
lost.  This effect is manifested through a shear-induced diffusion
which has been characterized through the mean square displacement
(MSD) of the particles~\cite{pine,brady-FP,acrivos}. The MSD
scales in the form: $\langle\Delta x^{2}\rangle \sim d^{2}
\dot{\gamma} t$, thus allows one to define an effective
diffusivity scaling like $D \sim d^{2} \dot{\gamma}$.

Characterizing the motion of the particles through the MSD clearly
suggests that a statistical description of their dynamics is
possible. Previously, this description was offered in
Ref.~\cite{brady-FP} by postulating a diffusion equation in which
the diffusivities have been constructed by analyzing the temporal
behavior of the position correlation function of the particles.
This approach allows the use of direct experimental measurements
or simulation results in order to describe particular
systems~\cite{bredveld}. Other theoretical and numerical studies
have characterized the relation between the transition to a
chaotic motion of the particles with the shear-induced diffusion
effect ~\cite{pine,brady-FP,acrivos,vulpiani}.

In this article, we offer a general description of this
shear-induced diffusion effect and the associated transition to
irreversibility which is based on the application of the second
law of thermodynamics, and on previous works devoted to analyze
the dynamics of a suspension of Brownian particles in the presence
of flows \cite{nosotros,njp}.

We calculate the entropy production of the system in the phase
space of the particles and find the corresponding Onsager
couplings~\cite{reviewMNET,nosotros,njp}. One of these couplings
is responsible for the dependence of the diffusion tensor on the
imposed velocity gradient, even in the limit of small specific
thermal energy: $k_BT/m \rightarrow 0$, with $k_BT$ the thermal
energy and $m$ the mass of a particle~\cite{nosotros}. This
dependence of the diffusion tensor on the velocity gradient leads
to the shear-induced diffusion effect that depends crucially on
hydrodynamic
interactions~\cite{zwanzig,mariano,acuna,NBrownies,saarloos}, and
the breaking of the fluctuation-dissipation relation (FDR) at
mesoscopic
level,~\cite{evans,nosotros,ryskin,leporini,drossinos,brady-PHYSA,hernandez,sengers}.

The article is organized as follows. In
Sec.~\ref{sec:fokker-planck} we use non-equilibrium thermodynamics
in phase space to formulate the mesoscopic model based on a
Fokker-Planck equation. Sec.~\textbf{III} is devoted to deriving a
Smoluchowski equation having an effective diffusion tensor
accounting for the shear-induced diffusion observed in
experiments. In section~\textbf{IV} we present Lattice-Boltzmann
simulations characterizing the shear-induced diffusion effect via
the power spectrum of particle movements affected by hydrodynamic
interactions. Finally, in Sec.~\textbf{V} we discuss our main
results.

\section{Mesoscopic entropy production for
the dynamics of a suspension in external flow}
\label{sec:fokker-planck}

We consider a suspension of $N$ non-interacting spherical
particles of radius $a$ and mass $m$ in a fluid which moves with
velocity $\vec{v}^{\,\,0}(\vec{r},t)$. Since the system is in
contact with a heat bath that evolves in time, it is necessary to
determine the physical nature of the coupling forces in order to
adequately describe its dynamics. This objective can be achieved
by taking into account two factors. The first one is that the
evolution of the system can be described at mesoscopic level by
means of the normalized $N$-particle probability distribution
function $P^{(N)}(\Gamma ^{N},t)$, that depends on the
instantaneous positions $\{\vec{r}\}^{N} \equiv
(\vec{r}_1,...,\vec{r}_N)$ of the particles and their velocities
$\{\vec{u}\}^{N} \equiv (\vec{u}_1,...,\vec{u}_N)$ through the
phase space vector $\Gamma
^{N}=(\{\vec{r}\}^{N},\{\vec{u}\}^{N})$. The second factor takes
into account the fact that the interactions between the system and
the heat bath involve dissipation. This suggests the use of the
nonequilibrium entropy $s(t)$ as a thermodynamic potential from
which the entropy production $\sigma(t)$ can be calculated, and
used to obtain the explicit expressions for the coupling forces
\cite{degroot}.

To proceed in systematical way, we will first notice that the
probability distribution function satisfies the conservation law
\begin{equation}
\label{balance probd} \frac{\partial P^{(N)}}{\partial t}+\sum
_{i=1}^{N}\vec{u}_{i}\cdot \nabla _{\vec{r}_{i}}P^{(N)}=-\sum
_{i=1}^{N}\frac{\partial }{\partial \vec{u}_{i}}\cdot
\vec{J}_{\vec{u}_{i}},
\end{equation}
where $ \nabla _{\vec{r}_{i}} $ represents the gradient operator
with respect to the position vector $ \vec{r}_{i} $ and
$\vec{J}_{\vec{u}_{i}} $ is a diffusion current defined in phase
space. Integration of Eq.~(\ref{balance probd}) over the phase
space coordinates $\Gamma ^{N} $, under the assumption that $
P^{(N)} $ and $\vec{J}_{\vec{u}_{i}} $ vanish at the boundaries,
leads to the continuity equation: $ {\partial \rho }/{\partial
t}=-\nabla\cdot (\rho\vec{v})$ in which the average density field
$\rho (\vec{r},t)$ of the suspended particles is defined
by~\cite{NBrownies}
\begin{equation}
\label{densidad} \rho (\vec{r},t)=m \int \sum
_{i=1}^{N}P^{(N)}(\Gamma ^{N},t)
\delta(\vec{r}_{i}-\vec{r})d\Gamma ^{N},
\end{equation}
and the mean velocity field $ \vec{v}(\vec{r},t) $ is
\begin{equation}
\label{flujo difusion} \vec{v}(\vec{r},t)=\frac{1}{\rho }m\int
\sum _{i=1}^{N}\vec{u}_{i}P^{(N)}(\Gamma ^{N},t)\delta(\vec{r}_{i}-\vec{r})d\Gamma ^{N}.
\end{equation}
Here $ d\Gamma ^{N}=d\{\vec{u}\}^{N}d\{\vec{r}\}^{N} $ is the
volume element in the phase space of the particles.

One of the purposes of this section is to derive explicit
expressions for the currents $\vec{J}_{\vec{u}_{i}}$ which at this
point implicitly contain the mentioned coupling forces between
system and bath. Once these expressions are obtained, the
evolution equation for $P^{(N)}$ can be written. As we have
mentioned, $\vec{J}_{\vec{u}_{i}} $ can be obtained from the
entropy production of the system which follows from the Gibbs
entropy postulate~\cite{NBrownies}
\begin{equation} \label{postuladogibbs}
\delta s(t)= -k_{B}\int \sum _{i=1}^{N}P^{(N)}\ln
\frac{P^{(N)}}{P_{l.eq.}^{(N)}}\delta(\vec{r}_{i}-\vec{r})d\Gamma ^{N},
\end{equation}
where $\delta s$ is the entropy change with respect to a local
equilibrium reference state characterized by the local equilibrium
distribution function
\begin{equation}
\label{distribucion eq.local}
P_{l.eq.}^{(N)}=e^{\frac{m}{k_{B}T}\left[\mu _{B}-\sum
_{i=1}^{N}\frac{1}{2}(\vec{u}_{i}-\vec{v}_i^{0})^{2}\right]}.
\end{equation}
Here $ \mu _{B} $ is the local equilibrium chemical potential per mass unit
and $\vec{v}_i^{\,\,0}=\vec{v}^{\,\,0}(\vec{r}_i,t)$.

Following the rules of mesoscopic nonequilibrium
thermodynamics~\cite{reviewMNET}, we take the time derivative of
Eq.~(\ref{postuladogibbs}) and use~(\ref{balance probd}), an
integration by parts over $\Gamma^{N}$-space assuming that the
fluxes vanish at the boundaries, leads to a balance equation for
the entropy $s$ in which the entropy production contains three
contributions: $\sigma=\sum_{j=1}^{3} \sigma_j$. The first
contribution is related to the diffusion process in
$\{\vec{u}\}$-space
\begin{eqnarray}
\label{sigma1}
\sigma_1  & = & -\frac{m}{T}\int \sum
_{i=1}^{N}\vec{J}_{\vec{u}_{i}}\cdot \frac{\partial \mu }{\partial
\vec{u}_{i}}\delta(\vec{r}_{i}-\vec{r})d\Gamma ^{N},
\end{eqnarray}
where the nonequilibrium chemical potential $\mu(\Gamma^N,t)$ is
given by
\begin{equation}\label{mu}
\mu(\Gamma^N,t)=\frac{k_BT}{m}\ln P^{(N)}+\frac{m}{2}\sum
_{i=1}^{N}(\vec{u}_{i}-\vec{v}_i^{0})^{2}.
\end{equation}
The second contribution comes from diffusion of particles
with respect to the mean velocity, with diffusion current
$\vec{J}_{i}=(\vec{u}_{i}-\vec{v}_i)P^{(N)}$:
\begin{eqnarray}
\label{sigma2} \sigma_2  & = &  -\frac{m}{2T}\int \sum
_{i=1}^{N}\vec{J}_{i}\cdot \nabla_{\vec{r}_{i}}
(\vec{u}_{i}-\vec{v}^0_i)^2\delta(\vec{r}_{i}-\vec{r})d\Gamma
^{N}.
\end{eqnarray}
The third contribution corresponds to diffusion with respect to
the flow velocity whose current is
$\vec{J}_{i}^0=(\vec{u}_{i}-\vec{v}^0_i)P^{(N)}$,
\begin{eqnarray}
\label{sigma3}
\sigma_3  & = &
 -\frac{m}{T}\int \sum _{i=1}^{N}\vec{J}_{i}^0\cdot
\vec{F}_{i}\delta(\vec{r}_{i}-\vec{r})d\Gamma ^{N},
\end{eqnarray}
where $\vec{F}_{i}=\partial \vec{v}^0_i /\partial t$
is a non-stationary force related with the variation of the fluid
velocity with time.

According to the second law of thermodynamics, the entropy
production of the system must be positive definite $\sigma > 0$
for irreversible process. To satisfy this condition,
nonequilibrium thermodynamics establishes linear relationships
between currents and forces~\cite{degroot}. In particular, for
$\vec{J}_{\vec{u}_{i}}$ we obtain
\begin{equation} \label{relaciones fen}
\vec{J}_{\vec{u}_{i}}=- \sum _{j=1}^{N}P^{(N)}\vec{\vec{\alpha
}}_{ij} \cdot \frac{\partial \mu }{\partial \vec{u}_{j}}- \sum
_{j=1}^{N}P^{(N)}\vec{\vec{\epsilon }}_{ij}
\cdot(\vec{u}_{j}-\vec{v}_j^{0})\cdot \nabla_{\vec{r}_{j}}
\vec{v}_j^{0} + \sum _{j=1}^{N}P^{(N)}\vec{\vec{\zeta}}_{ij} \cdot
\vec{F}_{j},
\end{equation}
where the tensors $\vec{\vec{\alpha }}_{ij}$, $\vec{\vec{\epsilon }}_{ij}$
and $\vec{\vec{\zeta}}_{ij}$ are related to the Onsager coefficients
$\vec{\vec{L}}_{u_{i}u_{j}}$, $\vec{\vec{L}}_{u_{i}r_{j}}$ and
$\vec{\vec{L}}_{u_{i}v_{j}}$ in the form~\cite{NBrownies}
\begin{equation}
\label{coef. fenom}
\vec{\vec{\alpha }}_{ij}={\vec{\vec{L}}_{u_{i}u_{j}}}/{TP^{(N)}},
\,\,\,\,\,\,\,\, \vec{\vec{\epsilon
}}_{ij}={\vec{\vec{L}}_{u_{i}r_{j}}}/{TP^{(N)}},
\,\,\,\,\,\,\,\,
\vec{\vec{\zeta}}_{ij}={\vec{\vec{L}}_{u_{i}v_{j}}}/{TP^{(N)}}.
\end{equation}
The Onsager coefficients obey Onsager's relations in which
time-reversal symmetry must also be applied to the external drive:
$\vec{\vec{L}}_{u_{i}r_{j}}=-\vec{\vec{L}}_{r_{i}u_{j}}
$,~\cite{dufty-rubi}. From Eq. (\ref{relaciones fen}) it follows
that the system of particles is coupled to the heat bath by means
of thermal and entropic forces (first term on the right hand side
of the equation) and mechanical forces (last two terms of the
equation).

Substituting Eq.~(\ref{relaciones fen}) into the continuity
equation for the probability~(\ref{balance probd}), we arrive at
the multivariate Fokker-Planck equation describing the evolution
of the $N$-particle distribution function
\begin{eqnarray}
\label{fokker-planck general}
 \frac{\partial P^{(N)}}{\partial t}+\sum _{i=1}^{N}\nabla _{\vec{r}_{i}}
 \cdot ( \vec{u}_{i} P^{(N)}) = \sum _{i,j=1}^{N}\frac{\partial }{\partial \vec{u}_{i}}\cdot
 \left\{\left[(\vec{u}_{j}-\vec{v}_j^{0})\cdot \vec{\vec{\beta}}_{ij} - \vec{\vec{\zeta}}_{ij} \cdot
 \vec{F}_{j}\right]P^{(N)}  +
 \frac{k_{B}T}{m}\vec{\vec{\alpha }}_{ij}\cdot
 \frac{\partial P^{(N)}}{\partial \vec{u}_{j}}\right\},
\end{eqnarray}
where we have used Eqs.~(\ref{mu}) and (\ref{relaciones fen})
assuming that the coefficients are symmetric tensors. Finally we
introduced the friction tensor $\vec{\vec{\beta}}_{ij}$ leading to
the relation~\cite{nosotros}
\begin{equation}
\label{def. tensor de friccion}
 \vec{\vec{\alpha }}_{ij} = \vec{\vec{\beta}}_{ij}- \vec{\vec{\epsilon }}_{ij}\cdot \nabla_{\vec{r}_{j}} \vec{v}_j^{0} .
\end{equation}

It is important to mention that the combination
$\vec{\vec{\epsilon}}_{ij}\cdot\nabla_{\vec{r}_{j}}
\vec{v}^0_{j}$, entering in Eqs.~(\ref{fokker-planck general})
and~(\ref{def. tensor de friccion}), implies that the
fluctuation-dissipation theorem (FDT) connecting the drift and
diffusion terms of the Fokker-Planck equation is no longer valid
due to the presence of the
shear~\cite{nosotros,leporini,ryskin,sengers}. This important
consequence following from Eq. (\ref{fokker-planck general}) is
related to the shear-induced diffusion effect, as we will show in
the next section. In the case of a diluted suspension, similar
results for the diffusion term of the generalized Fokker-Planck
equation have been obtained by means of the kinetic theory of
gases in Ref.~\cite{rosalio-dufty}.

The coefficient  $\vec{\vec{\epsilon }}_{ij}$ is related to the
force exerted on the surface of a particle moving through a fluid
under flow conditions. For a spherical particle,
$\vec{\vec{\epsilon }}_{ij}$ has been calculated explicitly in
terms of the generalized Fax\'en theorem in
Ref.~\cite{mazur-bedeaux}, and used in Ref.~\cite{njp} to obtain
$\vec{\vec{\epsilon}}_{ij}=\epsilon_{0}\vec{\vec{\tilde{\epsilon}}}_{ij}$
with
\begin{equation}\label{Epsilon-fax}
\epsilon_{0} =\frac{1}{6}\frac{m}{k_{B}T}a^{2}\beta _{0}^{2}
\left( 1+2a \, \alpha\right) ,
\end{equation}%
where $\beta_0=6\pi \eta a/m$ is the Stokes friction coefficient
per mass unit and $\eta$ the viscosity of the fluid, $\alpha=(-i
\omega/\nu)^{1/2}$ is the inverse viscous penetration length,
$\omega$ is the frequency and $\nu$ the corresponding kinematic
viscosity~\cite{mazur-bedeaux,njp}. The tensor
$\vec{\vec{\tilde{\epsilon}}}_{ij}$ is related to the friction
tensor $\vec{\vec{\beta}}_{ij}$ and obeys the relation
$\vec{\vec{\tilde{\epsilon}}}_{ii}=\vec{\vec{1}}$ with
$\vec{\vec{1}}$ the unit tensor. In Ref.~\cite{njp} it has been
shown that $\vec{\vec{\zeta}}_{ij}$ is related to inertial effects
due to the change in time of $\vec{v}_0$ and has the form
$\vec{\vec{\zeta}}_{ij}=\zeta\vec{\vec{1}}\delta_{ij}$, with
$\zeta={\rho _{p}}/{\rho _{f}}$, $\rho _{p}$ the density of the
particle and $\rho _{f}$ the density of the heat bath. From
Eq.~(\ref{Epsilon-fax}) it follows that $\epsilon_{0}$
incorporates finite-size effects on the dynamics of the system
through the surface term $a^2$ and frequency-dependent corrections
to the diffusion coefficient through $\alpha$.

 Eqs.~(\ref{fokker-planck general})-(\ref{Epsilon-fax})
imply that the diffusion coefficient in velocity space,
$\left({k_{B}T}/{m}\right)\vec{\vec{\alpha }}_{ij}$, does not
vanishes in the limit $k_BT/m \rightarrow 0$ and, therefore, also
imply that the non-thermal contribution to the diffusion
coefficient associated with the Onsager coefficient
$\epsilon_{ij}$ may have important consequences on the dynamics of
a non-Brownian suspension of particles. We will show these
consequences in the next section.

The friction tensors $\vec{\vec{\beta}}_{ij}$ are affected by the
hydrodynamic interactions among particles and their dependence can
be inferred from its relation with the mobility tensors
$\vec{\vec{\mu}}_{ij}$: $\vec{\vec{\beta}}_{ij}\cdot
\vec{\vec{\mu}}_{ij}=\vec{\vec{1}}\delta_{ij}$. At lower-order
approximation, the multipole expansion of $\vec{\vec{\mu}}_{ij}$
takes the form~\cite{saarloos}
\begin{equation} \label{mobility}
\vec{\vec{\mu}}_{ij} \simeq \beta_0^{-1}\vec{\vec{1}}\delta_{ij}
+\beta_0^{-1}\left[\frac{3}{4}\frac{a}{r_{ij}}\left(\vec{\vec{1}}
+\hat{r}_{ij}\hat{r}_{ij}\right)(1-\delta_{ij})-\frac{3}{4}\frac{a}{r_{ij_s}}\left(\vec{\vec{1}}
+\hat{r}_{ij_s}\hat{r}_{ij_s}\right)\right].
\end{equation}
Here $\hat{r}_{ij}$ and $\hat{r}_{ij_s}$ are the unit relative
vectors between particles and between particle $j$ and the wall.
$\vec{r}_{ij}=\vec{r}_i-\vec{r}_j$ is distance between particles
whereas $r_{ij_s}$ is the magnitude of the vector that points from
sphere $i$ to the mirror image with respect to a wall of sphere
$j$. For $i=j$, Eq.~(\ref{mobility}) reduces to well-known results
for the mobility of a particle in the presence of a wall:
$\mu=\beta_0^{-1}\left(1-B_1{a}/{l}\right)$, with $l$ its distance
to the wall. The coefficient $B_1$ may take different values
depending on the direction of the motion of the particle with
respect to the plane of the wall~\cite{happy}.

\section{Shear-induced diffusion}

We will analyze in this section
the diffusion regime occurring at times
$t\gg\beta_0^{-1}$. Since experiments and simulations
give the self-diffusion coefficient~\cite{pine,brady},
we will focus our description on the dynamics
of a single particle whose reduced distribution function $\rho_k
(\vec{r},t)=m\int P^{(N)}\delta(\vec{r}_{k}-\vec{r})d\Gamma ^{N}$
satisfies the continuity equation
\begin{equation} \label{cont-1part}
\frac{\partial \rho_k }{\partial t}=-\nabla \cdot (\rho_k \vec{v}_k),
\end{equation}
which follows by integrating Eq.~(\ref{fokker-planck general})
over the phase space of the particles and where
$\vec{v}_k(\vec{r},t)=m\rho_k^{-1}\int \vec{u}_{k}P^{(N)}\delta(\vec{r}_{k}-\vec{r})d\Gamma ^{N}$.

The Smoluchowski equation for $\rho_k$ can be derived after
calculating the explicit expression for $\rho_k \vec{v}_k$. This
task can be carried out by calculating the evolution equations for
the momentum field $\rho_k \vec{v}_k$ and for the pressure tensor
of the $k$-th particle, defined as
\begin{equation} \label{P-tensor}
\vec{\vec{\mathrm{P}}}_{k}(\vec{r},t)= m\int
(\vec{u}_k-\vec{v}_k)(\vec{u}_k-\vec{v}_k) P^{(N)} \delta(\vec{r}_{k}-\vec{r})d\Gamma ^{N}.
\end{equation}

Following the method indicated in Ref.~\cite{nosotros}, we
take the time derivative of the definition of $\rho_k (\vec{r},t)$,
use Eq.~(\ref{fokker-planck general}) in the result and
perform an integration by parts assuming that the currents vanish
at the boundaries. After rearranging terms we
arrive at the equation
\begin{equation}
\rho_k \frac{d_k}{dt}\vec{v}_k + \nabla \cdot
\!\!\,\,\vec{\vec{\mathrm{P}}}_{k}\!\!\,\, = - \int  \sum
_{i=1}^{N} \vec{\vec{\beta}}_{ki} \cdot
(\vec{v}_i^{(2)}-\vec{v}^0_i) \rho^{(2)} \delta(\vec{r}_{k}-\vec{r})d\vec{r}_kd\vec{r}_i +\rho \zeta \vec{F} ,
\label{momentum}
\end{equation}%
where we have used the expression
$\vec{\vec{\zeta}}_{ij}=\zeta\vec{\vec{1}}\delta_{ij}$ and defined
the convective derivative ${d_k}/{dt}=\partial/\partial t +
\vec{v}_k \cdot
\partial/\partial \vec{r}$, and the two-particle reduced
distribution function $\rho^{(2)}(\vec{r}_k,\vec{r}_i,t)= m\int
P^{(N)} d\Gamma^{N-2}_{ki}$. Here, $d\Gamma^{N-2}_{ki}$ the
phase-space volume element of the $N-2$ particles including the
measure associated to the velocities $\vec{u}_k$ and $\vec{u}_i$.
In $\vec{v}_i^{(2)}$ the superscript indicates a dependence on
$\vec{r}_k$ and $\vec{r}_i$. The right hand side of
Eq.~(\ref{momentum}) represents the total drag force exerted by
the fluid on the particles. The first term on the right hand side
is the friction force including the presence of hydrodynamic
interactions that modifies the local value of the velocity field
in terms of the distribution of particles in the system.

The evolution equation for the pressure tensor
$\!\!\,\,\vec{\vec{\mathrm{P}}}_{k}\!\!\,\,$ can be derived by
following a similar procedure:
\begin{eqnarray} \label{evolP}
\frac{d_k}{dt}\!\!\,\,\vec{\vec{\mathrm{P}}}_{k}\!\!\,\,+
2\left[\left(\beta_0 \vec{\vec{1}}+ \nabla
\vec{v}_k+\frac{1}{2}\nabla \cdot \vec{v}_k \vec{\vec{1}}\right)
\cdot \!\!\,\,\vec{\vec{\mathrm{P}}}_{k}\!\!\,\,\right]^{s} =
\frac{2k_{B}T}{m} \beta_{0} \rho_{k}\left[\vec{\vec{1}}-\frac{m}{6
k_{B}T}a^2\beta_{0}\left( 1+2a \, \alpha\right)\nabla\vec{v}^0_k\right]^{s}
\,\,\,\,\,\,\,\,\,\,\,\,\,\,\,\,\,\,\,\,\,\,\,\,\,\,\,\,\,\,\,\,\,
\\
\,\,\,\,\,\,\,\,\,\,\,
\,\,\,\,\,\,\,\,\,\,\,\,\,\,\,\,\,\,\,\,\,\,\,\,\,\,\,\,\,\,\,\,\,
\,\,\,\,\,\,\,\,\,\,\,\,\,\,\,\,\,\,\,\,\,\,\,\,
 - 2\left[ \int
\sum _{i=1,i\neq k}^{N} \vec{\vec{\beta}}_{ki}\cdot
(\vec{v}_k^{(2)} -\vec{v}^0_k) (\vec{v}_i^{(2)} -\vec{v}^0_i)
\rho^{(2)} \delta(\vec{r}_{k}-\vec{r})d\vec{r}_kd\vec{r}_i \right]^{s} ,
\nonumber
\end{eqnarray}%
where to obtain the second term on the right hand side, we have
used the expressions (\ref{def. tensor de friccion}) and
(\ref{Epsilon-fax}) for $i=j$. The upper symbol $s$ means the
symmetric part of a tensor. The last term on the right hand side
of this equation contains the contribution of the hydrodynamic
interactions to the pressure tensor of the particle $k$. This
contribution enters through the cross-correlation functions of the
local velocities indicating how hydrodynamic interactions modify
the stresses in the system. In order to obtain a closed expression
for the pressure tensor
$\!\!\,\,\vec{\vec{\mathrm{P}}}_{k}\!\!\,\,$, it thus becomes
necessary to calculate the evolution equation for the
cross-correlation function
\begin{equation} \label{z-tensor}
\vec{\vec{\mathrm{C}}}_{ki}(\vec{r},t)= \int
(\vec{u}_k-\vec{v}_k)(\vec{u}_i-\vec{v}_i)  P^{(N)}
\delta(\vec{r}_{k}-\vec{r})d\Gamma^{N}.
\end{equation}

The evolution equation for $\vec{\vec{\mathrm{C}}}_{ki}$ can be
obtained in a similar way as we did to derive Eq. (\ref{evolP}).
The result is
\begin{eqnarray}\label{evolZ}
\frac{d_k}{dt}\vec{\vec{\mathrm{C}}}_{ki} + 2\left[\left( \nabla
\vec{v}_k+\frac{1}{2}\nabla \cdot \vec{v}_k \vec{\vec{1}}\right)
 \cdot \vec{\vec{\mathrm{C}}}_{ki}\right]^{s}
= \frac{2k_{B}T}{m} \int \sum _{j=1}^{N}
\vec{\vec{\alpha}}^{\,s}_{kj} \delta_{ji}\rho^{(2)}
\delta(\vec{r}_{k}-\vec{r})d\vec{r}_kd\vec{r}_i
\,\,\,\,\,\,\,\,\,\,\,\,\,\,\,\,\,\,\,\,\,\,\,\,\,\,\,\,\,\,\,\,\,\,\,\,\,\,\,
\,\,\,\,\,\,\,\,\,\,\,\,\,\,\,\,\,\,\,\,\,
\\
\,\,\,\,\,\,\,\,\,\,\,\,\,\,\,\,\,\,\,\,\,\,\,\,\,\,\,\,\,\,\,\,
\,\,\,\,\,\,\,\,\,\,\,\,\,\,\,\,\,\,\,\,\,\,\,\,\,\,\,\,\,\,\,\,\,
\,\,\,\,\,\,\,\,\,\,\,\,\,\,\,\,\,\,\,\,\,\,\,\,
 - 2\left[ \int
\sum _{j=1}^{N} \vec{\vec{\beta}}_{kj}\cdot (\vec{v}_j^{(2)}
-\vec{v}^0_j) (\vec{v}_i^{(2)} -\vec{v}^0_i) \rho^{(2)}
\delta(\vec{r}_{k}-\vec{r})d\vec{r}_kd\vec{r}_i \right]^{\,s} ,
\nonumber
\end{eqnarray}%
where $k\neq i$. To derive Eqs.~(\ref{evolP}) and~(\ref{evolZ}),
we have neglected the contributions arising from higher order
moments of the time-ordered hierarchy since they relax faster than
the ones present in those equations~\cite{nosotros}.

At times $t\gg\beta_0^{-1}$, we can obtain from~(\ref{evolP}) the
following constitutive equation for
$\!\!\,\,\vec{\vec{\mathrm{P}}}_{k}\!\!\,\,$
\begin{eqnarray} \label{P-longTime}
\!\!\,\,\vec{\vec{\mathrm{P}}}_{k}\!\!\,\, \simeq
\frac{k_{B}T}{m} \rho_{k}\left[\vec{\vec{1}}-\frac{m}{6
k_{B}T}a^2\beta_{0}\left( 1+2a \, \alpha\right)\nabla\vec{v}^0_k\right]^{s}
\,\,\,\,\,\,\,\,\,\,\,\,\,\,\,\,\,\,\,\,\,\,\,\,\,\,\,\,\,\,\,\,\,\,\,\,\,\,
\,\,\,\,\,\,\,\,\,\,\,\,\,\,\,\,\,\,\,\,\,\,\,\,\,\,\,\,\,\,\,\,\,\,\,\,\,\,\,\,\,\,\,
\\
\,\,\,\,\,\,\,\,\,\,\,\,\,\,\,\,\,\,\,\,\,\,\,\,\,\,\,\,\,\,\,\,\,\,\,\,\,\,
\,\,\,\,\,\,\,\,\,\,\,\,\,\,\,\,\,\,\,\,\,\,\,\,\,\,\,\,\,\,\,\,\,\,\,\,\,\,\,\,\,\,\,
-\left[ \beta_{0}^{-1}\int
\sum _{i=1,i\neq k}^{N} \vec{\vec{\beta}}_{ki}\cdot
(\vec{v}_k^{(2)} -\vec{v}^0_k) (\vec{v}_i^{(2)} -\vec{v}^0_i)
\rho^{(2)} \delta(\vec{r}_{k}-\vec{r})d\vec{r}_kd\vec{r}_i \right]^{s}
, \nonumber
\end{eqnarray}%
where we have used that $(\nabla \vec{v}_k)_{ij}
\beta_{0}^{-1}\ll 1$, $\nabla \cdot \vec{v}_k \beta_{0}^{-1} \ll
1$, in accordance with the experiments~\cite{pine}.

For times $t\gg\beta_0^{-1}$, Eq. (\ref{evolZ}) can be rewritten
in similar form as Eq. (\ref{P-longTime}) by extracting the term
$j=k$ from the sum on the second term at the right-hand side,
multiplying the resulting relation by $\beta_0^{-1}
\vec{\vec{\beta}}_{ki}$ and performing the sum over $i$. One
obtains the expression
\begin{eqnarray}\label{alpha-betas}
\frac{k_{B}T}{m} \beta_0^{-1} \int \sum_{i=1,i\neq k}^{N}
\vec{\vec{\beta}}_{ki}\cdot\vec{\vec{\alpha}}_{ki} \rho^{(2)}
\delta(\vec{r}_{k}-\vec{r})d\vec{r}_kd\vec{r}_i
\,\,\,\,\,\,\,\,\,\,\,\,\,\,\,\,\,\,\,\,\,\,\,\,\,\,\,\,\,\,\,\,\,\,\,\,\,\,\,
\,\,\,\,\,\,\,\,\,\,\,\,\,\,\,\,\,\,\,\,\,
\\
\,\,\,\,\,\,\,\,\,\,\,\,\,\,\,\,\,\,\,\,\,\,\,\,\,\,\,\,\,\,\,\,
\,\,\,\,\,\,\,\,\,\,\,\,\,\,\,\,\,\,\,\,\,\,\,\,\,\,\,\,\,\,\,\,\,
\,\,\,\,\,\,\,\,\,\,\,\,\,\,\,\,\,\,\,\,\,\,\,\, \simeq \int \sum
_{i=1,i\neq k}^{N} \vec{\vec{\beta}}_{ki}\cdot (\vec{v}_k^{(2)}
-\vec{v}^0_k) (\vec{v}_i^{(2)} -\vec{v}^0_i) \rho^{(2)}
\delta(\vec{r}_{k}-\vec{r})d\vec{r}_kd\vec{r}_i ,\nonumber
\end{eqnarray}
where we have used $\vec{\vec{\beta}}_{kk}=\beta_0\vec{\vec{1}}$,
kept terms of the order $(a/r_{ij})^2$ and neglected terms of the
order $(a/r_{ij})^4$ and higher . This approximation is valid up
to intermediate volume fractions of the suspended particles.

\subsection{Effective medium approximation }

The Smoluchowski equation for $\rho_k$ can be obtained
from~(\ref{cont-1part}),~(\ref{momentum}), (\ref{P-longTime})
and~(\ref{alpha-betas}) by assuming an effective medium
approximation in which the test particle $k$ performs its motion
in a fluid incorporating the effects of hydrodynamic interactions
in average form~\cite{freed1}. In our description, this assumption
does not consider the possibility of direct collisions among
particles, and thus we expect that it is valid up to intermediate
volume fractions of the suspended particles. Operationally, this
approximation can be implemented by substituting the averages
appearing on the right-hand side of equations~(\ref{momentum}),
(\ref{P-longTime}) and~(\ref{alpha-betas}) by integrals over a
continuum variable. Thus, the last term on the right-hand side of
Eqs.~(\ref{P-longTime}) and (\ref{alpha-betas}) may be written as
\begin{equation}\label{CMT3}
\int
\sum_{i=1,i\neq k}^{N}  \vec{\vec{\beta}}_{ki}\cdot
(\vec{v}_k^{(2)} -\vec{v}^0_k) (\vec{v}_i^{(2)} -\vec{v}^0_i)
\rho^{(2)} \delta(\vec{r}_{k}-\vec{r})d\vec{r}_kd\vec{r}_i
\simeq
\int \vec{\vec{\beta}}(\vec{r}')\cdot\vec{\vec{\mathrm{C}}}^{\,*}(\vec{r}-\vec{r}',t)
d\vec{r}',
\end{equation}
where we have introduced the non-local velocity cross-correlation
function $\vec{\vec{\mathrm{C}}}^{\,*}(\vec{r}-\vec{r}',t)$. In
similar form, for the momentum field we have the relation
\begin{equation}\label{CMT1}
\int \sum _{i=1,i \neq k}^{N} \vec{\vec{\beta}}_{ki} \cdot
(\vec{v}_i^{(2)}-\vec{v}^0_i) \rho^{(2)}
\delta(\vec{r}_{k}-\vec{r})d\vec{r}_kd\vec{r}_i \simeq \int
\vec{\vec{\beta}}(\vec{r}') \cdot
\left(\vec{v}-\vec{v}^0\right)_{\vec{r},\vec{r}'}
\rho^{(2)}(\vec{r}-\vec{r}',t)d\vec{r}' ,
\end{equation}
where we have defined
$\left(\vec{v}-\vec{v}^0\right)_{\vec{r},\vec{r}'}\equiv
\vec{v}(\vec{r}-\vec{r}',t)-\vec{v}^0(\vec{r}-\vec{r}',t)$. Using
Eqs.~(\ref{def. tensor de friccion}) and (\ref{Epsilon-fax}), the
right-hand side of Eq.~(\ref{alpha-betas}) can be rewritten as
\begin{eqnarray}\label{CMT2}
\int \sum_{i=1,i\neq k}^{N}
\vec{\vec{\beta}}_{ki}\cdot\vec{\vec{\alpha}}^{\,s}_{ki}
\rho^{(2)} \delta(\vec{r}_{k}-\vec{r})d\vec{r}_kd\vec{r}_i \simeq
\int
\vec{\vec{\alpha}}^{\,*}(\vec{r}')\rho^{(2)}(\vec{r}-\vec{r}',t)
d\vec{r}' =
\,\,\,\,\,\,\,\,\,\,\,\,\,\,\,\,\,\,\,\,\,\,\,\,\,\,\,\,\,
\,\,\,\,\,\,\,\,\,\,\,\,\,\,\,\,\,\,\,\,\,\,\,\,\,\,\,\,\,\,\,  \\
\,\,\,\,\,\,\,\,\,\,\,\,\,\,\,\,\,\,\,\,
\int\vec{\vec{\beta}}(\vec{r}')\cdot\vec{\vec{\beta}}(\vec{r}')
\rho^{(2)}(\vec{r}-\vec{r}',t)d\vec{r}' - \frac{m}{6
k_{B}T}a^2\beta_{0}^2\left( 1+2a \, \alpha\right)
[\nabla\vec{v}^0]^{\dag} \cdot\int
[\vec{\vec{\beta}}(\vec{r}')\cdot\vec{\vec{\tilde{\epsilon}}}(\vec{r}')]^{\dag}\rho^{(2)}(\vec{r}-\vec{r}',t)
d\vec{r}', \nonumber
\end{eqnarray}
where the upper $\dag$ means the transpose of a tensor and we have
used the fact that $\nabla\vec{v}^0$ does not depends on position.
The tensor
$\vec{\vec{\alpha}}^{\,*}(\vec{r}')=\vec{\vec{\beta}}(\vec{r}')\cdot\vec{\vec{\alpha}}(\vec{r}')$
has been defined to simplify the notation in subsequent relations.
In the long-time limit the substitution of Eqs. (\ref{CMT3}) and
(\ref{CMT2}) into (\ref{P-longTime}) yields the constitutive
equation for the pressure tensor
\begin{equation}\label{P-constitut}
\!\!\,\,\vec{\vec{\mathrm{P}}}\!\!\,\, \simeq \frac{k_{B}T}{m}
 \left[\rho\left(\vec{\vec{1}}-\frac{m}{6
k_{B}T}a^2\beta_{0}\left( 1+2a \,
\alpha\right)\nabla\vec{v}^0\right)- \beta_0^{-2} \int
\vec{\vec{\alpha}}^{*}(\vec{r}')\rho^{(2)}(\vec{r}-\vec{r}',t)
d\vec{r}'\right]^{\,s},
\end{equation}%
where we have assumed that the second term on the left hand side
of Eq.~(\ref{evolZ}) may be neglected, see Eq.~(\ref{P-longTime}).

Using now Eq.~(\ref{CMT1}) in~(\ref{momentum}) and
taking the long-time limit $t\gg\beta_0^{-1}$, from the momentum equation
we obtain
\begin{equation}\label{momentum-effective}
\nabla \cdot
\!\!\,\,\vec{\vec{\mathrm{P}}}\!\!\,\, \simeq
- \int \vec{\vec{\beta}}(\vec{r}') \cdot
\left(\vec{v}-\vec{v}^0\right)_{\vec{r},\vec{r}'}  \rho^{(2)}(\vec{r}-\vec{r}',t)d\vec{r}'+\rho \zeta \vec{F} .
\end{equation}%
The explicit expression for the constitutive relation of
the diffusion current $\rho \vec{v}$ of the particles
follows by assuming that the
spatial variation of the velocity field is small enough in order
to make the expansion
\begin{equation}\label{taylor-momentum}
\vec{v}(\vec{r}-\vec{r}',t)-\vec{v}^0(\vec{r}-\vec{r}',t) \simeq
\, \vec{v}(\vec{r})\,- \vec{v}^0(\vec{r}) + O\left[\left(\nabla
\vec{v}^0\right)^2\right],
\end{equation}%
where we have used $\vec{v}\simeq \vec{v}^0 +O(\nabla \ln\rho)$,
\cite{nosotros}.

Factorizing the two-particle distribution function in the form
\cite{mayorgas}: $\rho^{(2)}(\vec{r}-\vec{r}',t)\simeq
\rho(\vec{r})g(\vec{r}-\vec{r}',t)$, with $g(\vec{r}-\vec{r}',t)$
the two-particle correlation function, we define the effective
quantities
\begin{equation}\label{eff-mobility}
\vec{\vec{B}}(\vec{r},t;\phi) = \beta_{0}^{-1}\int{
\vec{\vec{\beta}}(\vec{r}')\cdot\vec{\vec{\beta}}(\vec{r}\,')
g(\vec{r}-\vec{r}',t;\phi,T)
d\vec{r}\,'\,\,\,\,\,\,\text{and}\,\,\,\,\,\,}
\vec{\vec{E}}(\vec{r},t;\phi) = \beta_{0}^{-1} \int{
\vec{\vec{\beta}}(\vec{r}')\cdot\vec{\vec{\tilde{\epsilon}}}(\vec{r}\,')
g(\vec{r}-\vec{r}',t;\phi,T) d\vec{r}\,'},
\end{equation}
where we have taken into account the fact that the two-particle
correlation function $g$ may in general depend on the volume
fraction and the temperature, \cite{Hill}.

After using these results into Eqs.
(\ref{CMT2})-(\ref{momentum-effective}), we obtain the following
constitutive equation for the diffusion current
\begin{equation}\label{momentum-constitut}
\rho \vec{v} \simeq   \rho \vec{v}^0 - \frac{k_BT}{m}
\vec{\vec{B}}^{-1} \cdot \left(\nabla \cdot \vec{\vec{A}}\right)
\rho + \rho \zeta \vec{\vec{B}}^{-1} \cdot \vec{F} -
\vec{\vec{D}}(\vec{r},t)\cdot \nabla\rho,
\end{equation}%
where we have identified the effective diffusion tensor
\begin{equation}\label{eff-diff}
\vec{\vec{D}}(\vec{r},t) = {k_{B}T}/{m} \vec{\vec{\mu}} +
\frac{a^2}{6}\left( 1+2a \, \alpha\right)
\left[\left(\vec{\vec{1}}+\vec{\vec{\tilde{\mu}}}\right)\cdot(\vec{\vec{E}}-\vec{\vec{1}})\cdot
\nabla\vec{v}^0 \right]^{\,s}.
\end{equation}%
Here $\vec{\vec{\mu}}=\beta_0^{-1}\vec{\vec{\tilde{\mu}}}$ is the
effective mobility tensor and we have introduced the dimensionless
tensors
\begin{equation}\label{A}
\vec{\vec{\tilde{\mu}}}=\beta_{0}\vec{\vec{B}}^{-1}-\vec{\vec{1}}
\,\,\,\,\,\,\text{and}\,\,\,\,\,\,
\vec{\vec{A}}=\left(\vec{\vec{1}}-\beta_{0}^{-1}\vec{\vec{B}}+
\frac{m}{6k_BT}a^2\beta_0\left( 1+2a \,
\alpha\right)(\vec{\vec{E}}-\vec{\vec{1}})\cdot\nabla\vec{v}^0\right)^{\,s}.
\end{equation}
Eqs. (\ref{eff-mobility})-(\ref{eff-diff}) show that the transport
coefficients in the coarse-grained description contain
hydrodynamic interactions in effective form through the
configurationally averaged tensors $\vec{\vec{B}}$ and
$\vec{\vec{E}}$. When $\nabla\vec{v}^0=0$, then the diffusion
tensor reduces to: $\vec{\vec{D}}(\vec{r},t)= (k_BT/m)
\vec{\vec{\mu}}(\vec{r},t;\phi)$, that is, in equilibrium the
diffusion tensor depends on hydrodynamic interactions through the
mobility tensor $\vec{\vec{\mu}}$, as expected. The
coarse-graining performed in this section thus leads to
incorporate the contribution of hydrodynamic interactions in the
dissipation of the reduced system by modifying the diffusion
coefficient making it anisotropic and position and time dependent.
Substituting now Eq.~(\ref{momentum-constitut}) into
(\ref{cont-1part}), we finally obtain the Smoluchowski equation
\begin{equation} \label{Smoluch} \frac{\partial \rho }{\partial
t}=-\nabla \cdot \left[\rho \vec{v}_0 -\rho
\beta_0^{-1}\left(\vec{\vec{1}}+\vec{\vec{\tilde{\mu}}}\right)\cdot
\vec{f} -\rho \zeta
\beta_0^{-1}\left(\vec{\vec{1}}+\vec{\vec{\tilde{\mu}}}\right)\cdot
\vec{F} \right]+\nabla \cdot \left(\vec{\vec{D}}\cdot \nabla
\rho\right),
\end{equation}%
where we have defined the force $\vec{f}$ due to hydrodynamic
interactions as $\vec{f}=-(k_BT/m) \nabla \cdot \vec{\vec{A}}$.

Eqs.~(\ref{eff-diff})-(\ref{Smoluch}) constitute the main result
of this section. The \emph{effective} diffusion tensor
$\vec{\vec{D}}$ contains two contributions. The first one depends
on the thermal energy per mass unit ($k_BT/m$) and is therefore
related to Brownian motion whereas the second one does not depends
on thermal fluctuations.

In the limit of vanishing specific thermal energy, $k_{B}T/m
\rightarrow 0$, Eq. (\ref{eff-diff}) leads to
\begin{equation}\label{eff-diff-nonthermal}
\vec{\vec{D}} = \frac{a^2}{6}\left( 1+2a \, \alpha\right)
\left[\left(\vec{\vec{1}}+\vec{\vec{\tilde{\mu}}}\right)\cdot(\vec{\vec{E}}-\vec{\vec{1}})\cdot
\nabla\vec{v}^0 \right]^{\,s}.
\end{equation}%
This expression has the same scaling on particle diameter and
shear rate as that observed in experiments and Stokesian dynamics
simulations~\cite{pine,brady}. The presence of
$\vec{\vec{\tilde{\mu}}}(\vec{r},t;\phi)$ and
$\vec{\vec{E}}(\vec{r},t;\phi)$ implies that the shear-induced
diffusion is mediated by hydrodynamic interactions. We then
conclude that hydrodynamic interactions are responsible for the
randomization in the motion of the suspended particles when an
oscillatory strain is applied on the system.

\subsection{The mean square displacement}

We will assume that the fluid velocity
$\vec{v}_0(\vec{r},t)=\vec{r}\cdot\vec{\vec{\gamma}}(t)$ is
imposed along the $x$ direction with $\vec{\vec{\gamma}}(t)$ the
time-dependent shear rate whose only non-vanishing component is
$\gamma_{yx}=\dot{\gamma}\,cos(\omega t)$. The shear rate is
related to the applied strain $\gamma_0$ by
$\dot{\gamma}={\gamma_0}\omega$. For convenience, we will assume
that inertial effects are negligible and that effective mobility
$\vec{\vec{\tilde{\mu}}} $ and $\vec{\vec{E}}$ do not depend on
time and position. This hypothesis is valid when the distribution
of the suspended particles does not changes significatively, that
is when $g(\vec{r}\,' -\vec{r},t;\phi)\sim g(\vec{r}\,';\phi)$.

The MSD of particle`s position vector can
be calculated by taking the time derivative of the expression
\begin{equation} \label{msd}
\langle r^2\rangle=\int(x^2+y^2)\,\rho\, d\vec{r},
\end{equation}
where we have considered the two-dimensional case. Substitution of
Eq.~(\ref{Smoluch}) into the result and an integration by parts
leads to
\begin{equation} \label{evolMSD}
\frac{d }{d t}\langle r^2\rangle= 2 \dot{\gamma}\, cos(\omega t)\,
\langle xy\rangle(t)+ 2 Tr[\vec{\vec{D}}] ,
\end{equation}
where $\langle xy\rangle(t)=\int xy\, \rho\, d\vec{r}$. In similar
form, we must derive the evolution equations for $\langle
xy\rangle(t)$, $\langle x^2\rangle(t)$ and $\langle
y^2\rangle(t)$. After solving the obtained set of differential
equations, for low shear rates and frequencies ($\dot{\gamma} <1$,
$\omega<1$) we may expand the MSD in a power
series of $\dot{\gamma}$ and $\omega$ to obtain
\begin{equation} \label{MSD1}
\langle r^2\rangle \simeq 4 D_0
\left[\tilde{\mu}_{xx}+\tilde{\mu}_{yy}\right]t+
\frac{1}{6}\tilde{\mu}_{xy}(E-1)d^2\dot{\gamma}t\left[1 +
\frac{24}{d^2 (E-1)}D_0t\right],
\end{equation}
where $D_0={k_{B}T}/{m\beta_{0}}$ is the one-particle diffusion
coefficient and $d=2a$ having assumed
$\tilde{\mu}_{xy}=\tilde{\mu}_{yx}$ and
$\vec{\vec{E}}=E\vec{\vec{1}}$. In the limit $k_BT/m \rightarrow
0$ when the particles are non-Brownian we obtain
\begin{equation} \label{MSD2}
\langle r^2\rangle \sim
\frac{1}{6}\tilde{\mu}_{xy}(\phi)[E(\phi)-1]d^2 \dot{\gamma} t.
\end{equation}
When Eq. (\ref{MSD2}) is expressed in terms of the number of
oscillations $n$ of the imposed flow, with $t=2\pi n/\omega$, it
gives: $\langle r^2\rangle \sim \frac{\pi}{3}\tilde{\mu}_{xy}(E-1)
d^2 \gamma_0 n$. This relation shows that hydrodynamic interaction
introduce a volume fraction dependence of the MSD of the particle,
thus giving an explanation based on thermodynamic arguments for
the scaling relation obtained in the experiments~\cite{pine}.

In the limit of massive particles, from Eq. (\ref{Smoluch}) it is
also possible to derive the evolution equation for the average
position of the particle defined through $\vec{R}(t)=\int \vec{r}
\rho d\vec{r}$. By taking the time derivative of this definition
and integrating by parts one obtains
\begin{equation} \label{evol-R}
\frac{d }{d t}\vec{R}(t)=\vec{R}(t)\cdot\nabla\vec{v}_0(t)- \zeta
\beta_0^{-1}\vec{\vec{G}}_1(\vec{R};\phi)\cdot\frac{d }{d
t}\nabla\vec{v}_0(t)+\frac{a^2}{6}\vec{\vec{G}}_2(\vec{R};\phi)
\cdot \nabla\vec{v}_0(t),
\end{equation}%
where we have introduced the quantities $\vec{\vec{G}}_1=\langle
\left(\vec{\vec{1}}+\vec{\vec{\tilde{\mu}}}\right)\cdot \vec{r}
\rangle $ and $\vec{\vec{G}}_2=\langle\nabla \cdot
[\left(\vec{\vec{1}}+\vec{\vec{\tilde{\mu}}}\right)
\cdot(\vec{\vec{E}}-\vec{\vec{1}})]
\rangle $ and the bracket indicates an average over $\rho$. Eq.
(\ref{evol-R}) is a nonlinear equation for $\vec{R}(t)$ in which
the nonlinearities are a consequence of hydrodynamic interactions
through the terms $\vec{\vec{G}}_1$ and $\vec{\vec{G}}_2$. In a
first approximation, the last term at the right hand side of Eq.
(\ref{evol-R}) establishes that hydrodynamic interactions become
significant when the Reynolds number defined by $Re \equiv
dv_0/\nu$ satisfies the condition
\begin{equation} \label{Re}
Re > \frac{24 d^2 \omega}{\nu}h(\phi),
\end{equation}%
where $\nu$ is the kinematic viscosity of the heat bath and the
function $h(\phi)$ takes into account that $(G_2)_{ij}$ is a
function of $\phi$. In obtaining this relation we have neglected
the second term at the right hand side of (\ref{evol-R}) since
$\beta_0^{-1}$ is a very small quantity and scaled time with
$\omega$ and lengths with $d$.

Eqs. (\ref{MSD2}) and (\ref{Re}) show that the transition to the
irreversibility is mediated by hydrodynamic interactions that
introduce a dependence on volume fraction of the shear-induced
diffusion coefficient.

\section{Shear-induced diffusion from Lattice-Boltzmann simulations}

In this section, we analyze the shear-induced diffusion effect by
means of Lattice-Boltzmann simulations. This method allows us to
study the dependence of the effective diffusion coefficient as a
function of the relevant parameters of the problem, the Reynolds
number Re and the volume fraction $\phi$. The power spectrum of
the components of the trajectories of the particles is used to
show that the randomization of particle movements is due to an
increasing number of modes produced by hydrodynamic interactions
as Re and $\phi$ increase.

We use the two-dimensional model D2Q9 for the Lattice-Boltzmann
method with the BGK approximation~\cite{qian92,bgk54}. In this
model, the space is discretized in a two dimensional square
lattice with nine velocities ($\mathbf c_i$,  $i=0 \ldots 8$)
allowed. The particle distribution functions $f(\mathbf r,t)$, at
site $r$ and time $t$ evolve according to the equation
\begin{equation}
f_i(\mathbf r + \mathbf c_i) -f_i(\mathbf r,t) = -\frac{1}{\tau}
\left[ f_i(\mathbf r,t) - f^{(eq)}_i(\mathbf r,t) \right],
\end{equation}
where $\tau$ is the dimensionless relaxation time related to
viscosity and $f_i^{(eq)}$ are the local equilibrium distribution
functions,
\begin{equation}
f_i^{(eq)} = w_i \rho \left[ 1 + 3\mathbf c_i \cdot \mathbf u + \frac{9}{2} (\mathbf c_i \cdot \mathbf u )^2
- \frac{3}{2} u^2 \right].
\end{equation}
In this equation, $w_i=4/9$, $1/9$, $1/36$ are the weights
associated to the lattice~\cite{he97} for each set of velocities
$|\mathbf c_i| = 0,1,\sqrt 2$ and $\rho$  and $\mathbf u$ are the
density and velocity defined by
\begin{equation}
\rho (\mathbf r,t) = \sum_i f_i (\mathbf r,t),\qquad
\mathbf u (\mathbf r,t) = \frac{1}{\rho} \sum_i f_i (\mathbf r,t) \mathbf c_i.
\end{equation}
The viscosity is related to the dimensionless relaxation time by
$\nu = c_s^2 (\tau - 1/2)$, where $c_s = 1/\sqrt 3$ is the speed
of sound in the D2Q9 model.

The no-slip boundary conditions are simulated on the solid
particles and the torques and forces are also evaluated to update
the particles position at all times ~\cite{aidun98}. The
interactions among particles are implemented with the method
proposed in Ref.~\cite{ladd94} and with the corrections proposed
in Ref. ~\cite{aidun98}. The walls of the cavity use the
bounce-back boundary condition, which consists in reversing the
incoming particle distribution function after the stream process.


The numerical simulations are carried out in a cavity of
$H^\ast=11.33 $ and $W^\ast = 44.66$ where the dimensions were
scaled with the radius of the particle. The relaxation time and
the radius of the particles are kept constant in all simulation at
$\tau =20.0$, $r=4.5$ and $f^\ast = 10.0$ and we have varied the
Reynolds number and the volume fraction. The dimensionless
frequency was scaled with the magnitude of the shear rate.

\begin{figure}
\includegraphics{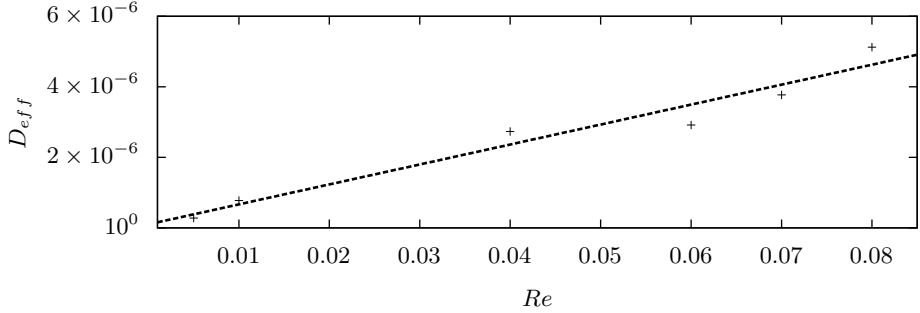}
\caption{\label{fig:Phi14e-2-freq10} Effective diffusion
coefficient as a function of Re (symbols) for a fixed particle
concentration of $\phi = 0.14$ and $f^\ast = 10.0$. The solid line
is a linear fit with slope of $5.68\times 10^{-5}$.
}
\end{figure}

In the first set of numerical simulations, we fixed the particle
concentration $\phi = 0.14$ and the dimensionless frequency
$f^*=10$, varied the Reynolds number and determined the effective
diffusion coefficient $D$ as shown in
Fig.~\ref{fig:Phi14e-2-freq10}. A linear dependence of $D_{eff}$
on Re has been obtained to a good approximation, as expected from
experiments and theoretical results. Using the definition
$Re=d^2\dot{\gamma}/\nu$, from Eq. (\ref{MSD2}) we may also obtain
the linear relation for the effective diffusivity as a function of
the Reynolds number: $D_{eff}=\tilde{\mu}_{xy}(\phi)[E(\phi)-1]\nu
Re/6$. The data obtained from simulations can be used to give a
rough estimate of the magnitude of hydrodynamic interactions in
terms of the parameter $\tilde{\mu}_{xy}(\phi)[E(\phi)-1]$. Given
$d=9$ and $\nu=11.25$ we obtain
$\tilde{\mu}_{xy}(\phi)[E(\phi)-1]\sim 10^{-4}$.  This small value
indicates that 3D Lattice-Boltzmann simulations are required in
order to do a quantitative comparison with experiments.

\begin{figure}
\includegraphics{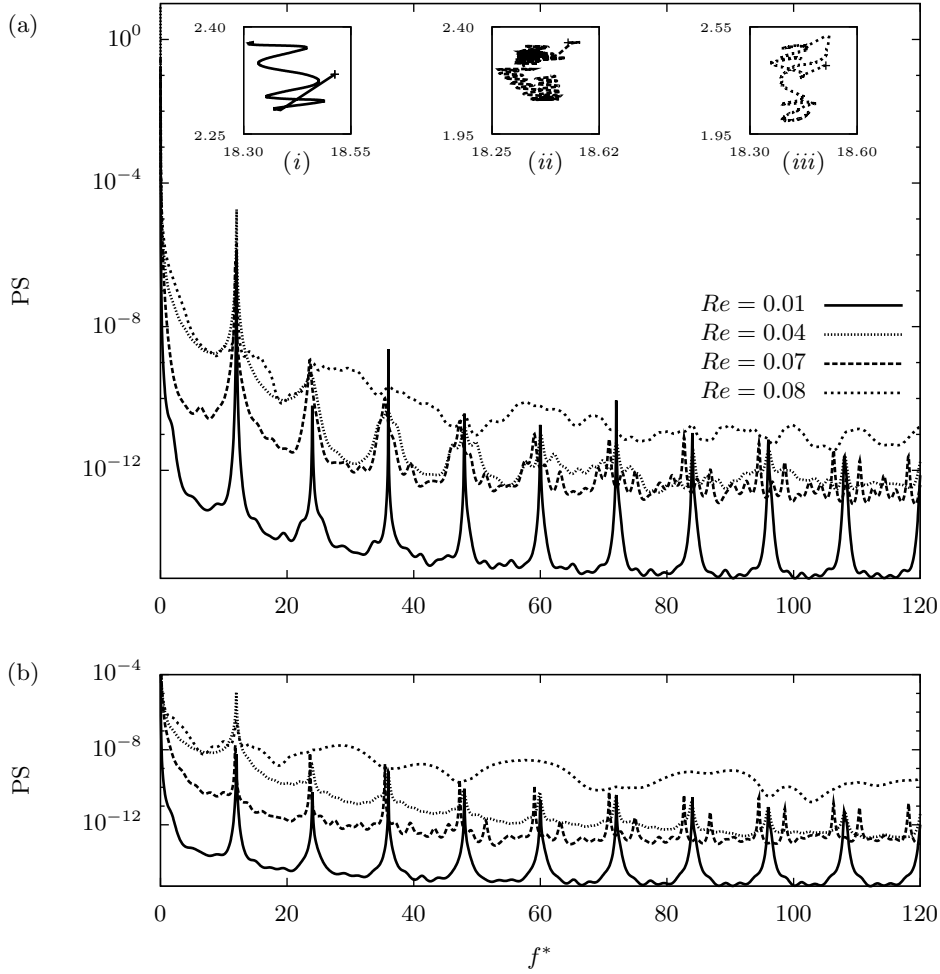}
\caption{\label{fig:ps-Re}
Power spectrum for the same particle at different Reynolds numbers for (a) the x-movement and (b) the y-movement.
The cross in the insets indicates the initial position of the particle, the same in the three cases.
Notice that maximum displacement increases with increasing Reynolds number.
}
\end{figure}

\begin{figure}
\includegraphics{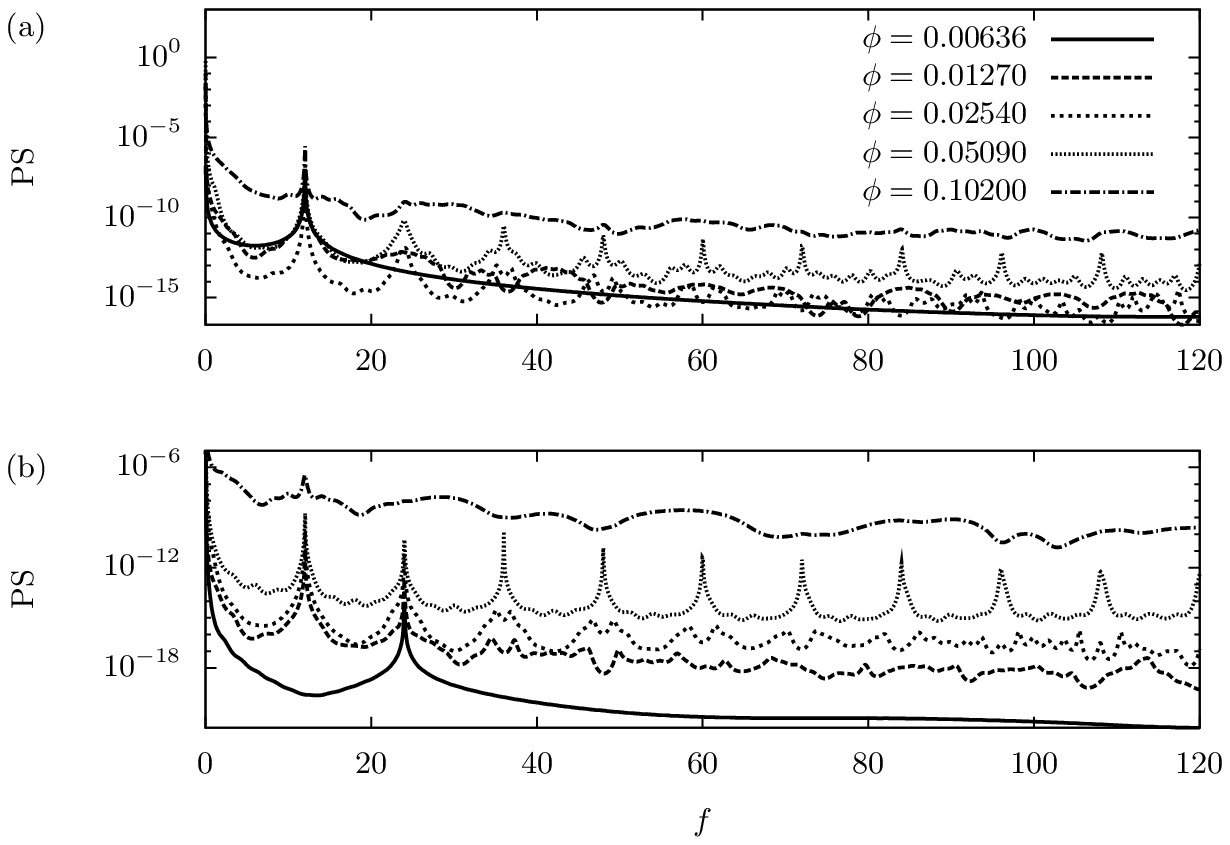}
\caption{\label{fig:ps-f-Re8-phi} } Power spectrum for different
values of $\phi$ at $Re = 0.08$ for a) the $x$ and b) $y-$movement
of the trajectory of a particle.
\end{figure}

In order to better discern the mechanisms leading to the
shear-induced diffusion, we calculated the power spectrum (PS) of
the components of the trajectories for fixed $\phi$ and Re, as it
can be respectively seen from Figs.~\ref{fig:ps-Re}
and~\ref{fig:ps-f-Re8-phi}.

In Fig.~\ref{fig:ps-Re} (a), we present the PS of $x(t)$ for four
different values of Re and $\phi=0.14$. The insets represent three
different trajectories for the same particle with the same initial
condition for $Re =0.01$, $Re =0.07$ and $Re =0.08$. For lowest
Reynolds number (solid line), the PS presents more pronounced
peaks located at the excitation frequency and its harmonics. The
trajectory in the inset (\emph{i}) (solid line) corresponds to
this spectrum and shows a very regular behavior. From the PS for
the case of $Re = 0.04$ (dotted line) it follows that the dynamics
in the $x$-component keeps the main peak and harmonics at the same
position than in the previous case, but small peaks start to
appear between the harmonics. This is a consequence of the
hydrodynamic interactions between the particles that give rise to
new frequencies in the dynamics of the system. In the case $Re =
0.07$ (dashed line) there is a shift of the harmonics and the new
frequencies are better defined. The corresponding trajectory
(dashed line) is shown in the inset (\emph{ii}). Finally, for  $Re
= 0.08$ (short-dashed line) the harmonics disappear and the energy
is more homogeneously distributed for frequencies larger than the
one imposed. The corresponding trajectory (dotted line), shown in
the inset (\emph{iii}), is irregular.  In Fig.~\ref{fig:ps-Re} (b)
we present the PS corresponding to the $y-$movement of the
particle for different Re at a fixed $\phi$. From this set of PS
we can appreciate that both the $x$ and $y-$movements are coupled
with the exciting frequency, as well as the fact that the
harmonics have a small shift and the energy is distributed in more
frequencies. The PS of $x(t)$ and $y(t)$ shown in Figs.
~\ref{fig:ps-Re} (a) and (b) agree with theoretical results in two
ways. First, they indicate a coupling between different modes,
like in Eq. (\ref{evolMSD}). Second, the power spectra reflects
that  hydrodynamic interactions become important only for Reynolds
numbers larger than a certain values, as established by Eq.
(\ref{Re}).

The theoretical results (\ref{eff-diff-nonthermal}) and
(\ref{MSD2}) indicate that the shear-induced effect is caused by
hydrodynamic interactions and, although it is not explicitly
shown, one then expects that the effect also depends on particle
concentration. This was confirmed by performing another set of
simulations keeping $Re=0.08$ constant and varying $\phi$ to
analyze the influence of hydrodynamic interactions on the dynamics
of the particles through the PS of $x(t)$ and $y(t)$. The results
are shown in Figs.~\ref{fig:ps-f-Re8-phi}(a) and (b),
respectively. For the lower volume fraction $\phi = 0.00636$ (one
particle) the PS for $x(t)$, Fig.~\ref{fig:ps-f-Re8-phi}(a), has
only one peak at the exciting frequency. This result is expected
from theory, Eq. (\ref{evol-R}), since for sufficiently small
strains and frequencies the leading contribution corresponds to
the first term on the right hand side, which is proportional to
the applied strain. For the $y(t)$,
Fig.~\ref{fig:ps-f-Re8-phi}(b), the peak is shifted to the right
from the exciting frequency, showing a weak coupling between the
$x$ and $y$ motions. For $\phi = 0.01270$ (two particles), $x(t)$
presents harmonics which disappear at frequencies much larger than
the exciting one whereas for $y(t)$ the peak at the exciting
frequency appears with an incipient presence of harmonics,
implying that hydrodynamic interactions are weak. For $\phi =
0.02540$ (four particles) three harmonics can be identified for
both $x(t)$ and $y(t)$. For $\phi = 0.05090$ (eight particles ) a
larger number of harmonics can be identified in $x(t)$ with about
half of the energy in comparison with the exciting frequency. For
the $y(t)$, the harmonics are clearly identified and have the same
energy as the exciting frequency. New frequencies arise between
the harmonics implying that hydrodynamic interactions introduce
new modes in the dynamics of the particles. This result is
expected from the evolution equation (\ref{evol-R}) which predicts
that for strains or particle concentrations larger than a certain
critical value, new modes will appear in the behavior of
$\vec{R}(t)$ due to the contributions of the nonlinear terms.
Finally, for $\phi = 0.102$ (16 particles), the only peak for the
$x(t)$ and $y(t)$ is located at the exciting frequency and the
energy is now homogeneously distributed in more frequencies. These
results indicate that an increase in particle concentration
enhances the effects of hydrodynamic interactions which in turn
are responsible for distributing the energy in a growing number of
modes. They also reflect the fact that hydrodynamic interactions
also introduce a dependence on volume fraction through the
dependence on the volume fraction of the coefficients
$\tilde{\mu}_{xy}(\phi)$ and $E(\phi)$.

\section{Conclusions}

In this paper, we have analyzed the shear-induced diffusion effect
in suspensions under oscillatory shear by means of a thermokinetic
theory based on the calculation of the entropy production at
mesoscopic level. We have found that an Onsager coupling between
thermal and non-thermal effects containing hydrodynamic
interactions is responsible for this effect.

By calculating the entropy production of the $N$-particle system
in contact with a heat bath and identifying the corresponding
forces and currents, we have derived a Fokker-Planck equation for
the $N$-particle phase space distribution function having an
effective diffusion coefficient in which the coupling between
thermal and non-thermal effects breaks the validity of the
fluctuation-dissipation theorem. After contracting the description
over velocity space, assuming that the diffusion regime is well
established and using an effective medium approximation, we
derived a Smoluchowski equation for the single particle
distribution function containing an effective diffusion tensor
incorporating the cross-effects of the imposed flow and
hydrodynamic interactions, Eq. (\ref{eff-diff}). In the limit
$k_BT/m \rightarrow 0$, this diffusion tensor yields the same
scaling on particle diameter and applied strain as found in
experiments and simulations, $D\simeq d^2
\tilde{\mu}_{xy}(\phi)[E(\phi)-1] \gamma_0/6 $. Our analysis may
also explain the influence thermal noise on this effect.

The Lattice-Boltzmann simulations we performed were used to show
in more detail the dependence of the shear-induced diffusion on
the Reynolds number and the particle volume fraction. To this
effect we have used the power spectrum of the components of
particle trajectories. As expected from theory, we found a linear
dependence of the diffusion coefficient $D$ as a function of the
Reynolds number Re. When increasing Re for fixed $\phi$ as well as
$\phi$ for fixed Re, we obtained that the power spectrum shows an
increasing contribution of new modes, period doubling and finally,
for large Re and $\phi$, a loss of characteristic frequencies
indicating a stochastic behavior induced by hydrodynamic
interactions, in accordance with theoretical results expressed
through the mean square displacement (\ref{MSD2}). It is important
to emphasize that the 2D simulations performed clearly indicate
that up to intermediate volume fractions the hydrodynamic
interactions are responsible for the shear-induced diffusion
effect, in agreement with the theoretical prediction. A more
precise description could be carried out from 3D simulations in
order to perform a quantitative comparison with experiments.

The expression for the shear-induced diffusion coefficient in Eq.
(\ref{MSD2}) and the condition (\ref{Re}) show that the transition
to the irreversibility is due to hydrodynamic interactions and
that it depends on the volume fraction, in accordance with
experiments and theoretical simulations. Therefore, our study
gives a theoretical explanation of the shear induced effect and
the transition to the irreversibility based on the pertinent
analysis of the entropy production at mesoscopic level.

\section{ACKNOWLEDGMENTS}

We acknowledge Prof. D. J. Pine for interesting discussions on
experiments, Dr. R. Rechtman for valuable commentaries on the
numerical simulations and Dr. G. Ruiz Chavarr\'ia for technical
support. GBV acknowledges financial support by
DGAPA-UNAM and ISH to grant DGAPA-IN102609.


\begin{thebibliography}{99}

\bibitem{pine} D. J. Pine, J. P. Gollub, J. F. Brady, A. M.
Leshansky, Nature {\bf 438}, 997 (2005).

\bibitem{acrivos} D. Drazer, J. Koplik, B. Khusid, A. Acrivos,
J. Fluid Mech. {\bf 460}, 307 (2002).

\bibitem{bredveld} V. Breedveld, D. van den Ende,
A. Tripathi, A. Acrivos, J. Fluid Mech.  {\bf 375}, 297-318 (1998).

\bibitem{taylor} G. I. Taylor, J. Friedman, \emph{Low Reynolds Number
Flows} (National Committe on Fluid Mechanics Films, Encyclopedia
Britannica Educational Corp., United States, 1996).

\bibitem{brady-FP} A. Seriou, J. F. Brady, J. Fluid Mech. {\bf 506 }
285 (2004).

\bibitem{vulpiani} G.
Boffeta, M. Cencini, M. Falcioni, A. Vulpiani, Phys. Rep. {\bf
356}, 367 (2002).

\bibitem {nosotros} I. Santamar\'{\i}a-Holek,
D. Reguera and J. M. Rubi,  Phys. Rev. E {\bf63}, 051106 (2001).

\bibitem {njp} I. Santamar\'{\i}a-Holek, J. M. Rubi,
A. P\'erez-Madrid,  New J. Phys {\bf 7}, 35 (2005).

\bibitem {reviewMNET}D. Reguera, J. M. G. Vilar, J. M. Rubi,
J. Phys. Chem. B {\bf 109}, 21502 (2005).

\bibitem {zwanzig} R. Zwanzig, Adv. Chem. Phys. {\bf 15}, 325 (1969).

\bibitem {saarloos} W. V. Saarloos, P. Mazur, Physica A
{\bf 120}, 77-102 (1983)and Physica A {\bf 127}, 451-472 (1984).

\bibitem {mariano} M. L\'opez de Haro, J. M. Rubi, J.Chem. Phys.
{\bf 88}, 1248 (1987).

\bibitem {acuna}L. Yeomans-Reyna, H. Acu\~na-Campa,
M. Medina-Noyola, Phys. Rev. E {\bf 62}, 3395 (2000).

\bibitem {NBrownies} J. M. Rubi and P. Mazur, Physica A
{\bf 250}, 253 (1998).

\bibitem {ryskin} G. Ryskin,  Phys. Rev. Lett. {\bf 61} 01442
(1988).

\bibitem {leporini} R. Mauri, D. Leporini, Europhys. Lett.,
{\bf 76} 1022–1028 (2006).

\bibitem {evans} S. Sarman, D. J. Evans, A. Baranyai,  Phys. Rev.
A {\bf 46}, 893 (1992).

\bibitem {drossinos} Y. Drossinos, M. W. Reeks,  Phys. Rev. E
{\bf 71}, 031113 (2005).

\bibitem {brady-PHYSA} G. Subramanian, J. F. Brady, Physica A
{\bf 334}, 343 (2004).

\bibitem {hernandez} A. V. Popov, R. Hernandez J. Chem. Phys.
{\bf 126}, 244506 (2007).

\bibitem{rosalio-dufty} R. Rodr\'iguez, E. Salinas-Rodr\'iguez,
J. Dufty, J. Stat. Phys. 32, 279 (1983).

\bibitem {sengers} J. V. Sengers, J. M. Ort\'iz de Z\'arate,
J. Non-Equilib. Thermodyn. {\bf 32}, 319–329 (2007).

\bibitem {degroot} S. R. de Groot, P. Mazur,
\emph{Non-equilibrium Thermodynamics }, (Dover, New York, 1984).

\bibitem {dufty-rubi} J. W. Dufty and J. M. Rubi , Phys. Rev, A
{\bf36}, 222 (1987).

\bibitem {mazur-bedeaux} P. Mazur and D. Bedeaux, Physica A
{\bf 76}, 235 (1974).

\bibitem {happy} J. Happel and H. Brenner,
\emph{ Low Reynolds number hydrodynamics } (Kluwer Academic
Publishers, Dordrecht, 1991).

\bibitem {brady} G. Bossis and J. F. Brady, J. Chem. Phys
{\bf 91}, 1866 (1989).

\bibitem {freed1} K. F. Freed and M. Muthukumar, J. Chem. Phys.
{\bf 69}, 2657 (1978).

\bibitem {mayorgas} M. Mayorga, L. Romero-Salazar and J. M. Rubi,
Physica A {\bf 307}, 297 (2002).

\bibitem {Hill} T. L. Hill, \emph{An introduction to Statistical
Thermodynamics} (Dover, New York, 1986).

\bibitem {qian92} Y.~Qian, D.~d'Humieres and P.~Lallemand,
Eur. Phys. Lett. {\bf 17}, 479 (1992).

\bibitem {bgk54} P.~L.~Bhatnagar, E.~P.~Gross and M.~Krook,
Phys. Rev. {\bf 94}, 511 (1954).

\bibitem {he97} X. He and L.S. Luo, Phys. Rev. E
{\bf 56}, 6811 (1997).

\bibitem {aidun98} C.K. Aidun, Y. Lu and E.J. Ding,
J. Fluid Mech. {\bf 373}, 287 (1998).

\bibitem {ladd94} A. J. C. Ladd, J. Fluid Mech. {\bf 271}, 285
(1994); A. J. C. Ladd, J. Fluid Mech. {\bf 271}, 311 (1994).

\end{thebibliography}
\end{document}